\documentclass[conference]{IEEEtran}
\IEEEoverridecommandlockouts
\usepackage{cite}
\usepackage{amsmath,amssymb,amsfonts,mathtools}
\usepackage{graphicx}
\usepackage{textcomp}
\usepackage{xcolor}
\def\BibTeX{{\rm B\kern-.05em{\sc i\kern-.025em b}\kern-.08em
    T\kern-.1667em\lower.7ex\hbox{E}\kern-.125emX}}

\usepackage{bm}
\usepackage{comment}

\usepackage{tikzpeople}
\usetikzlibrary{calc}
\usetikzlibrary{decorations.pathreplacing,decorations.markings,shapes.geometric}
\usetikzlibrary{external}
\usepackage{algorithm}
\usepackage[noend]{algpseudocode}
\usepackage{algorithmicx}

\tikzset{naming/.style={align=center,font=\small}}
\tikzset{antenna/.style={insert path={-- coordinate (ant#1) ++(0,0.25) -- +(135:0.25) + (0,0) -- +(45:0.25)}}}
\tikzset{station/.style={naming,draw,shape=dart,shape border rotate=90, minimum width=10mm, minimum height=10mm,outer sep=0pt,inner sep=3pt}}
\tikzset{mobile/.style={naming,draw,shape=rectangle,minimum width=12mm,minimum height=6mm, outer sep=0pt,inner sep=3pt}}
\tikzset{radiation/.style={{decorate,decoration={expanding waves,angle=90,segment length=4pt}}}}

\tikzset{axis/.style={thick,-latex}}
\tikzset{vec/.style={thick,blue}}
\tikzset{univec/.style={thick,red,-latex}}
\usetikzlibrary{patterns}
\usepackage{siunitx}
\DeclareSIUnit[quantity-product = ]\percent{\char`\%}
\usepackage{booktabs}
\usepackage{dirtytalk}
\begin{document}

\title{Distributed Algorithm for Cooperative Joint Localization and Tracking Using Multiple-Input Multiple-Output Radars 
\thanks{$^*$ These authors contributed equally. \\
This work is funded by the Thomas B. Thriges Foundation grant 7538-1806.}
}

\author{
    \IEEEauthorblockN{Astrid Holm Filtenborg Kitchen$^*$, Mikkel Sebastian Lundsgaard Brøndt$^*$, Marie Saugstrup Jensen$^*$, Troels Pedersen,\\ and Anders Malthe Westerkam
    }
    \IEEEauthorblockA{Aalborg University, Aalborg, Denmark. Email: \{troels, amw\}@es.aau.dk}     
}

\maketitle
\begin{abstract}
We propose a distributed joint localization and tracking algorithm using a message passing framework, for multiple-input multiple-output radars. We employ the mean field approach to derive an iterative algorithm. The obtained algorithm features a small communication overhead that scales linearly with the number of radars in the system. The proposed algorithm shows good estimation accuracy in two simulated scenarios even below \qty{0}{dB} signal to noise ratio. In both cases the ground truth falls within the \qty{95}{\percent} confidence interval of the estimated posterior for the majority of the track.
\end{abstract}
\section{Introduction}
\noindent In recent years, drones have become more and more present in public and private air spaces due to their advancing technology. 
When used with malicious intent, they raise concerns regarding public, national, and private safety \cite{Miantezila2022Sssb}. Drones can breach restricted air spaces, e.g. at airports where a drone-plane collision could have fatal consequences \cite{Shvetsova2021Esas}. Despite regulations prohibiting drone operations in certain areas, numerous instances of these laws being breached continue to occur \cite{FTREPORTERS2018Dfco}.
As a result, security systems must be able to detect unauthorized drone activity to respond to these intrusions. 

There is a wide variety of drone detection and tracking methods relying on sensors such as radars or cameras with varying motion and measurement models, followed by efficient estimation models that handles uncertainties \cite{ZitarRaedAbu2023IRoD}. Of these sensors, radars are one of the most effective for aerial target surveillance as they operate in all-day, all-weather conditions \cite{clemente2021radar}, whereas vision-based systems suffer in partial occlusion conditions \cite{YuQian2008OTaR, UzairMuhammad2021BVEf}. 
Drones are generally small, high maneuvering, and fly with low altitude and velocity compared to large aircraft. 
In relation to radars, drones have a low radar cross section (RCS) and quickly changing kinematic parameters, which makes drones challenging to detect and track accurately with conventional radar systems. 
To increase the time on target, multiple-input multiple-output (MIMO) radar systems can be employed as they illuminate the entire scene of interest continuously with the trade-off being the power density leading to low signal to noise ratio (SNR) conditions. Hence, there is a pressing need for reliable methods to detect and track targets in low SNR conditions for MIMO radars.

Different approaches have been taken to enhance tracking performance in low SNR conditions using MIMO radars \cite{HuangDarong2023SASI, FangXin2023EAED, 8249147,Anders}. A commonly used method for tracking moving objects are different variations of Kalman filters (KFs). Additionally, KFs are being incorporated with machine learning models to enhance the performance and add more adaptability to the tracking process
\cite{ZitarRaedAbu2023IRoD, HaarnojaTuomas2017BKLD}.
In \cite{ZaraiKhaireddine2023Imtc}, the extended KF was replaced with an adaptive Monte Carlo method and combined with the joint probabilistic data association filter (JPDAF) for multi-target tracking. The JPDAF determines target presence by thresholding, making it susceptible to missed detections and false alarms, especially in cluttered environments. As an alternative, track-before-detect (TBD) algorithms have access to all radar measurements. Several articles have proposed TBD algorithms, based on recursive Bayesian theory. In \cite{HuangDarong2023SASI} a sequential Monte Carlo method was presented, jointly detecting
and tracking a target with constant velocity in \qty{-10}{dB} SNR using a 4D MIMO radar, whereas \cite{Anders} uses the mean field approach for Bayesian localization and tracking (BLaT) in \qty{-1}{dB} with a MIMO radar. 
In \cite{LehmannF.2012RBFf}, the evolution of the probability of a present target was tracked for each range-Doppler cell, but the complexity of this approach is tied to the specific radar settings, as well as relying on a discretized target vector. 
The contributions \cite{LaiYangming2023JDaL, YiWei2020SLCJ} use bistatic MIMO radars with widely separated antennas for joint multi-target detection and localization.
By separating the antennas, a target can be observed from multiple angles, enhancing spatial diversity \cite{FishlerE.2006SDiR}.
To enhance performance at low SNR, raw measurements of all receivers are directly sent to a fusion center and jointly processed \cite{RuixinNiu2012TLaT, GodrichHana2010TLAG,ZhangGuoxin2023DTLW,YangShixing2022MDSf}. However, this approach incurs a higher signaling overhead and increased computational cost.

In this paper, we extend the BLaT algorithm from \cite{Anders} to support multiple communicating MIMO radars and incorporate backwards smoothing with a recursive Bayesian filter.  
The resulting multiple radar BLaT (MRBLaT) algorithm, estimates target parameters in Cartesian coordinates, allowing the  posteriors from each MIMO radar to be combined in a global coordinate system.
The performance is evaluated by using simulations of multiple 3$\times$3 MIMO radars using time division multiplexing (TDM) and is compared to a multi input KF with backwards smoothing utilizing all available data and without thresholding. By using a network of MIMO radars, this approach improves the performance of jointly localizing and tracking a single target in a low SNR environment.

\begin{figure}
    \centering
    \includegraphics{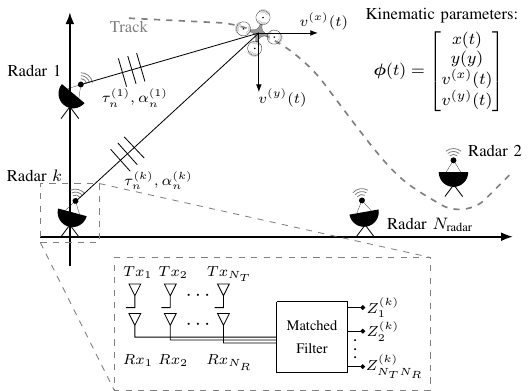}
    \caption{Overview of the scenario under consideration with $N_{\text{radar}}$ radars are placed throughout the scene. Each radar consists of $N_T$ transmitters transmitting orthogonal signals, that are reflected by the target which is governed by its kinematic parameters and collected by $N_R$ receivers. At each radar the collected signals are matched filtered resulting in $N_T N_R$ outputs.}
    \label{fig:Drone_detection}
\end{figure}
\section{Signal Model}
\noindent Following  Fig.~\ref{fig:Drone_detection}, we consider a single target in a clutter-free environment. The target can be described by its kinematic parameters: $\boldsymbol{\phi}(t) = [x(t), y(t), v^{(x)}\!(t), v^{(y)}\!(t)]^\top$, where $(x(t), y(t))$ denote the position of the target in a global Cartesian coordinate system, while $v^{(x)}\!(t)$ and $v^{(y)}\!(t)$ denote the velocity along each of the coordinate axes. The target is illuminated by $N_{\text{radar}}$ monostatic MIMO radars each with $N_T$ transmitting antennas and $N_R$ receiving antennas. All antennas are considered isotropic. Transmitter $m$ on radar $k$ denoted as $m^{(k)}$ emits a signal $\text{Re}\{u^{\left(m^{(k)}\right)}\!(t)\mathrm{e}^{i\omega_c t}\}$, where $u^{\left(m^{(k)}\right)}\!(t)$ denotes the complex baseband signal with angular carrier frequency $\omega_c=2\pi f_c$ and $i$ the imaginary unit. For each transmitter the transmitted signals are assumed mutually orthogonal. We assume no mutual interference, which could be achieved by e.g. frequency division of the radars. For simplicity, each radar has the same transmission scheme resulting in $u^{\left(m^{(k)}\right)}\!(t) = u^{(m)}\!(t)$. We consider the transmission of a MIMO pulse as the transmission from all $N_T$ transmitters, resulting in a total transmission time of $T_{Tx}N_{T}$. The transmission repeats with an interval $\Delta t$, i.e. the MIMO radars have a pulse repetition frequency (PRF) of $1/{\Delta t}$. Each MIMO pulse is reflected by the target and collected by the receiver on the radar from which it originated. It is further assumed that the target is slow moving w.r.t. $T_{Tx}N_{T}$, such that the \say{stop-and-hop} approach may be employed, and that the RCS of the target is constant across time and radars. The kinematic parameters are considered time invariant between pulses, i.e., $\boldsymbol{\phi}(t) = \boldsymbol{\phi}_n$ for $n\Delta t \leq t \leq (n+1)\Delta t$. The Doppler shift is ignored due to the low velocity. 

Finally, assuming only a direct path and the target being in the far-field of all radars, the signal received at each receiver~$j$ on radar $k$ after baseband conversion can be modeled as
\begin{multline}\label{eq:signalmodel}
    \hspace{-3.4mm} \mathcal{Y}^{(j,k)}_n(t) = \overbrace{\sum_{m=1}^{N_T} \alpha\left(\tau^{(k)}_n\right) A^{(j,m,k)}\!(x_n,y_n) u^{(m)}\!\left(t-\tau^{(k)}_n\right)}^{\Tilde{s}_n^{(j,k)}\!(t)} \\+ w^{(j,k)}_n\!(t),
\end{multline}
where $w^{(j,k)}_n\!(t)$ is complex circularly symmetric white Gaussian noise with zero mean and variance $\sigma_w^2$, $A^{(j,m,k)}$ is the steering matrix elements, and $\tau^{(k)}_n$ is the two way time delay between radar $k$ and the target. The path loss $\alpha(\tau_n^{(k)})$ is calculated using the radar range equation. 

The received signal is sampled and matched filtered which provides $N_TN_R$ complex signal vectors of length $N_s$, which is the number of samples. In the frequency domain the signal at radar $k$ reads
\begin{equation}\label{eq:freqdomain}
    \bm{Z}_n^{(k)} = \overbrace{\Tilde{\bm{S}}_n \otimes \bm{U}^{*}}^{\bm{S}_n(\phi_n)} + \Tilde{\bm{W}}_n^{(k)}\in \mathbb{C}^{N_RN_T\times N_s},
\end{equation}
with $\Tilde{\boldsymbol{S}}_n \in \mathbb{C}^{N_R\times N_s}$, $\boldsymbol{U}\in \mathbb{C}^{N_T\times N_s} $ being the collection of samples into a matrix, $\otimes$ being the column-wise Kronecker product, and $*$ the complex conjugate. The post match filtered noise $\Tilde{\bm{W}}_n^{(k)}$ is colored circular Gaussian noise with zero mean and precision matrix $\boldsymbol{\Lambda}_Z$ independent for each virtual element.

\section{Bayesian Network} \label{sec:VMP}
\begin{figure}
    \centering
    \includegraphics{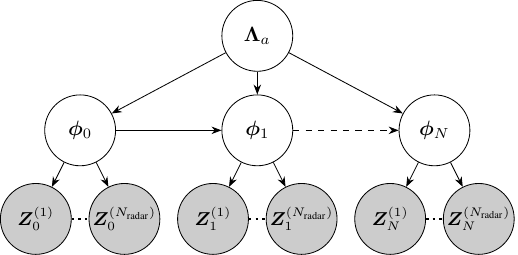}
    \caption{Variational framework where each node represents a stochastic variable. The shaded nodes represent observed variables.}
    \label{fig:Bayesian_network}
\end{figure}

\noindent The problem at hand is to estimate the kinematic parameters of the target at time $n$, denoted as $\boldsymbol{\phi}_n$, from the received signals at all radars $\{\boldsymbol{Z}_j^{(\forall N_{\text{radar}})}\}_{j\leq n}$. The estimation algorithm is derived based on the Bayesian network in Fig.~\ref{fig:Bayesian_network}. The evolution of the track is governed by the underlying kinematics. Assuming linear motion during $N_{Tx}T_{Tx}$ we use a Markov model,
\begin{align}\label{eq:kenimatic_model}
    \bm{\phi}_n = \bm{T}\bm{\phi}_{n-1} + \bm{G}\bm{a}, \quad \bm{a}|\bm{\Lambda}_a \sim \mathcal{N}(\bm{a};\bm{0},\bm{\Lambda}_a).
\end{align}
Here $\bm{T}$ denotes the kinematic matrix in Cartesian coordinates, while $\bm{G}$ denotes the process noise matrix, where
\begin{align}
    \bm{T} = 
    \begin{bmatrix}
    1 & 0 & \Delta t & 0 \\
    0 & 1 & 0 & \Delta t \\
    0 & 0 & 1 & 0 \\
    0 & 0 & 0 & 1
    \end{bmatrix}\!\!, \ 
    \bm{G} =
    \begin{bmatrix}
    \frac{\Delta t^2}{2} & 0 & 0 & 0 \\ 
    0 & \frac{\Delta t^2}{2} & 0 & 0\\
    0 & 0 & \Delta t & 0 \\
    0 & 0 & 0 & \Delta t
    \end{bmatrix}\!\!.
\end{align}
The joint distribution thus reads,
    \begin{multline}
    \hspace{-3.3mm} p\!\left(\left\{\bm{Z}^{(k)}_0,\dots,\bm{Z}^{(k)}_n\right\},\bm{\phi}_0,\dots,\bm{\phi}_n,\bm{\Lambda}_a\right)= p(\bm{\phi}_0\vert \bm{\Lambda}_a) p(\bm{\Lambda}_a)\\ \times \prod_{n=1}^{N}\left(\prod_{k=1}^{N_{\text{radar}}} p\left(\bm{Z}^{(k)}_n\vert \bm{\phi}_n\right)\!\right) p(\bm{\phi}_n\vert \bm{\phi}_{n-1},\bm{\Lambda}_a) .
    \end{multline}
Note that the prior $p(\bm{\phi}_0|\bm{\Lambda}_a)$ is complicated by the kinematic process \eqref{eq:kenimatic_model} as any stationary solution will have the variance of $p(\bm{\phi}_0|\bm{\Lambda}_a)$ tending to infinity. For this reason we use an improper distribution $p(\bm{\phi}_0|\bm{\Lambda}_a) = 1$. 

\section{Variational Message Passing}
\noindent To estimate $\{\bm{\phi}_n\}$ we seek the posterior, $p(\{\bm{\phi}_n\}|\{\bm{Z}\},\bm{\Lambda}_a)$, however this is intractable. In lieu, we employ the mean field approach and approximate it by a surrogate function,
 \begin{equation}
    q(\bm{\phi}_0,\hdots,\bm{\phi}_{N},\bm{\Lambda}_a) = q(\bm{\Lambda}_a)\prod_{n = 0}^N q(\bm{\phi}_n),
 \end{equation}
obtained by minimizing the Kullback-Leibler (KL) divergence w.r.t. the true posterior. Following \cite{bishop2007}, the surrogates $q(\boldsymbol{\phi}_n)$ and $q(\boldsymbol{\Lambda}_a)$ can be expressed as
\begin{multline}\label{eq:surrogate_for_phi}
    \hspace{-4mm} \ln(q(\bm{\phi}_n)) = \sum_{k=1}^{N_{\text{radars}}}\left(\ln(p(\bm{Z}_n^{(k)})\vert \bm{\phi}_n)\right)  \\ +\mathbb{E}_{\backslash \bm{\phi}_n}[\ln(p(\bm{\phi}_n\vert \bm{\phi}_{n-1},\bm{\Lambda}_a))] + \mathbb{E}_{\backslash \bm{\phi}_n}[\ln(p(\bm{\phi}_{n+1}|\bm{\phi}_n,\bm{\Lambda}_a))] \\+ \text{const.}
\end{multline}
\begin{multline}\label{eq:surrogate_for_Lambda}
    \hspace{-4mm} \ln q(\bm{\Lambda}_a) = \sum_{n=1}^N\mathbb{E}_{\backslash \bm{\Lambda}_a}[\ln(p(\bm{\phi}_n|\bm{\phi}_{n-1},\bm{\Lambda}_a))] \\ + \ln(p(\bm{\Lambda}_a)) + \text{const.}
\end{multline}
The surrogates are calculated recursively using a message passing scheme. Each term in \eqref{eq:surrogate_for_phi} and \eqref{eq:surrogate_for_Lambda} can be viewed as a message passed from the nodes connected to the node under consideration, i.e., the nodes in the neighborhood denoted $\mathcal{N}_{\text{node}}$.  
The messages are in the following calculated one by one in the order they appear in \eqref{eq:surrogate_for_phi} and \eqref{eq:surrogate_for_Lambda}.

The message $\epsilon^{(\bm{Z}^{(k)}_n\to \bm{\phi}_n)}$ can be expanded as,
\begin{multline}\label{eq:Data_message}
    \ln\!\left(p\!\left(\bm{Z}^{(k)}_n\vert \bm{\phi}_n\right)\!\right)\! \propto \\ -\!\left\langle\! \bm{S}_n^{(k)}\!(\bm{\phi}_n) - \bm{Z}_n^{(k)}\vert \bm{\Lambda}_{Z}\vert \bm{S}_n^{(k)}\!(\bm{\phi}_n) - \bm{Z}_n^{(k)}\!\right\rangle.
\end{multline}
Here $\langle\cdot|\cdot\rangle$ is the bra-ket notation for inner products. The function in \eqref{eq:Data_message} is not recognized as a known distribution in $\bm{\phi}_n$. Instead, we approximate this message by a Gaussian,
\begin{equation}
    \epsilon^{(\bm{Z}_n^{(k)}\rightarrow\bm{\phi}_n)} = \mathcal{N}\!\left(\bm{\phi}; \bar{\bm{\epsilon}}^{(\bm{Z}_n^{(k)}\rightarrow\bm{\phi}_n)},\left(\bar{\bar{\bm{\epsilon}}}^{(\bm{Z}_n^{(k)}\rightarrow\bm{\phi}_n)}\right)^{-1}\right)\!,
\end{equation}
with mean $\bar{\bm{\epsilon}}^{(\bm{Z}_n^{(k)}\rightarrow\bm{\phi}_n)}$ and covariance matrix $\bar{\bar{\bm{\epsilon}}}^{(\bm{Z}_n^{(k)}\rightarrow\bm{\phi}_n)}$. The optimal approximation of the message is found by minimizing the KL divergence between $p(\bm{Z}_n^{(k)}|\bm{\phi}_n)$ and a Gaussian in $\bm{\phi}_n$ as
\begin{align}\label{eq:minimize}
    \left\{\widehat{\overline{{\bm{\epsilon}}}}_n^{(k)},\widehat{\overline{\overline{{\bm{\epsilon}}}}}_n^{(k)}\right\} = \underset{\overline{{\bm{\epsilon}}},\overline{\overline{{\bm{\epsilon}}}}}{\arg\min}\ D_{KL}\!\left(\mathcal{N}\!\left(\bm{\phi}_n;\overline{\bm{\epsilon}},\overline{\overline{\bm{\epsilon}}}^{-1}\right)\!\Big\Vert p\!\left(\!\bm{Z}_n^{(k)}|\bm{\phi}_n\right)\!\right)\!,
\end{align}
where the superscript has been dropped for brevity. The KL divergence reads
\begin{multline}\label{eq:KL_div}
    \hspace{-3.3mm} D_{KL} = -\int_{\bm{\phi}_n}\mathcal{N}\!\left(\bm{\phi}_n;\overline{\bm{\epsilon}},\overline{\overline{\bm{\epsilon}}}^{-1}\right)\ln\!\left(p\!\left(\bm{Z}_n^{(k)}|\bm{\phi}_n\right)\!\right) \text{d}\bm{\phi}_n \\
    + \int_{\bm{\phi}_n}\mathcal{N}\!\left(\bm{\phi}_n;\overline{\bm{\epsilon}},\overline{\overline{\bm{\epsilon}}}^{-1}\right)\ln\!\left(\mathcal{N}\!\left(\bm{\phi}_n;\overline{\bm{\epsilon}},\overline{\overline{\bm{\epsilon}}}^{-1}\right)\!\right)\text{d}\bm{\phi}_n
\end{multline}
where the second term is the entropy of a Gaussian distribution, $\zeta(\bar{\boldsymbol{\epsilon}},\bar{\bar{\boldsymbol{\epsilon}}})$.
The first term is the expectation of $\ln(p(\bm{Z}_n^{(k)}|\bm{\phi}_n))$ w.r.t. a Gaussian and reads
\begin{align}\label{eq:expectation_term}
\begin{multlined}
    \hspace{-1.8mm} \mathbb{E}\!\left[-\!\left\langle \bm{S}_n^{(k)}\!(\bm{\phi}_n) - \bm{Z}_n^{(k)}\big\vert \bm{\Lambda}_{Z}\big\vert \bm{S}_n^{(k)}\!(\bm{\phi}_n) - \bm{Z}_n^{(k)}\right\rangle\!\right] \\ \propto -\!\left\langle \mathbb{E}\!\left[\bm{S}_n^{(k)}\!(\bm{\phi}_n)\right] - \bm{Z}_n^{(k)}\big\vert \bm{\Lambda}_{Z}\big\vert \mathbb{E}\!\left[\mathbf{S}_n^{(k)}\!(\bm{\phi}_n)\right] - \bm{Z}_n^{(k)}\right\rangle \\ - \text{Tr}\!\left(\mathbb{E}\!\left[\!\big\vert\bm{S}_n^{(k)}\!(\bm{\phi}_n)\big\rangle\big\langle\bm{S}_n^{(k)}(\bm{\phi}_n)\big\vert\right]\bm{\Lambda}_{Z}\!\right)\!.
\end{multlined}
\end{align}
These expectations are intractable, but can be approximated by the Delta method as
\begin{equation}
    \mathbb{E}\!\left[\bm{S}_n^{(k)}\!(\bm{\phi}_n)\right]\approx \bm{S}_n^{(k)}\!(\Bar{\bm{\epsilon}}),
\end{equation}
    \begin{multline}
    \mathbb{E}\!\left[\!\big|\bm{S}_n^{(k)}\!(\bm{\phi})\big\rangle\big\langle\bm{S}_n^{(k)}\!(\bm{\phi})\big|\right] \approx \big|\bm{S}_n^{(k)}\!(\Bar{\bm{\epsilon}})\big\rangle\big\langle\bm{S}_n^{(k)}\!(\Bar{\bm{\epsilon}})\big| \\ + \big|\nabla_\phi \bm{S}_n^{(k)}\!(\Bar{\bm{\epsilon}}_n)\big\rangle\Bar{\Bar{\bm{\epsilon}}}\big\langle\nabla_\phi \bm{S}_n^{(k)}\!(\Bar{\bm{\epsilon}}_n)\big|,
    \end{multline}
where $\nabla_\phi \bm{S}_n^{(k)}\!(\Bar{\bm{\epsilon}})$ is the Jacobian of $\mathbf{S}_n^{(k)}$ evaluated in $\Bar{\bm{\epsilon}}$. Inserting this in \eqref{eq:KL_div} gives,
\begin{align}\label{eq:KL_div_real}
    \begin{multlined}
        \hspace{-3.3mm} D_{KL} \propto -2\text{Re}\left\{\!\big\langle\bm{S}_n^{(k)}\!(\bar{\bm{\epsilon}}_n)\big|\bm{\Lambda}_{Z}\big|\bm{Z}_n^{(k)}\big\rangle\!\right\} \\+\big\langle\bm{S}_n^{(k)}\!(\Bar{\bm{\epsilon}}_n)\big|\bm{\Lambda}_{Z}\big|\bm{S}_n^{(k)}\!(\Bar{\bm{\epsilon}}_n)\big\rangle \\+ \text{Tr}\left(\Bar{\Bar{\bm{\epsilon}}}_n\big\langle \nabla_\phi \bm{S}_n^{(k)}\!(\Bar{\bm{\epsilon}}_n)\big|\bm{\Lambda}_{Z}\big|\nabla_\phi \bm{S}_n^{(k)}\!(\Bar{\bm{\epsilon}}_n)\big\rangle\right) - \zeta(\bar{\bar{\bm{\epsilon}}}).
    \end{multlined}
\end{align}
Equation \eqref{eq:KL_div_real} can then be minimized in $\bar{\bm{\epsilon}}$ and $\bar{\bar{\bm{\epsilon}}}$.

A problem arises with the path loss $\alpha$ as it contains information about the reflectivity of the target. Hence, $\alpha$ will be estimated using a maximum likelihood estimate using the previous mean of the kinematics,
\begin{equation}\label{eq:alpha_est}
    \hat{\alpha}^{(k)} = \frac{\big\langle\bm{S}_n^{(k)}\!\big(\bar{\bm{\phi}}_{n-1}\big)\big|\bm{\Lambda}_{Z}\big|\bm{Z}_n^{(k)}\big\rangle}{\big\langle\bm{S}_n^{(k)}\!(\bar{\bm{\phi}}_{n-1})\big|\bm{\Lambda}_{Z}\big|\bm{S}_n^{(k)}\big(\bar{\bm{\phi}}_{n-1}\big)\big\rangle}.
\end{equation}
Lastly the real operator in \eqref{eq:KL_div_real} causes instability when the optimization is performed numerically, due to the phase of $\alpha$ being proportional to $\omega_c\sqrt{x^2+y^2}$. Hence, any estimation error arising from \eqref{eq:alpha_est} will be overlaid as a fast oscillating cosine. It was found that discarding this phase instability by replacing the operator by the absolute value greatly improved estimation accuracy, hence the objective function used in the algorithm is,
\begin{align}\label{eq:KL_div_abs}
    \begin{multlined}
        \hspace{-3mm} D_{KL} \propto -\left|\hat{\alpha}^{(k)}\big\langle\Tilde{\bm{S}}_n^{(k)}\!(\bar{\bm{\epsilon}}_n)\big|\bm{\Lambda}_{Z}\big|\bm{Z}_n^{(k)}\big\rangle\right| \\+\big|\hat{\alpha}^{(k)}\big|^2\big\langle\Tilde{\bm{S}}_n^{(k)}\!(\Bar{\bm{\epsilon}})\big|\bm{\Lambda}_{Z}\big|\Tilde{\bm{S}}_n^{(k)}\!(\Bar{\bm{\epsilon}})\big\rangle \\\hspace{-3.2mm} + \big|\hat{\alpha}^{(k)}\big|^2\text{Tr}\!\left(\Bar{\Bar{\bm{\epsilon}}}\big\langle \nabla_\phi \Tilde{\bm{S}}_n^{(k)}\!(\Bar{\bm{\epsilon}})\big|\bm{\Lambda}_{Z}\big|\nabla_\phi \Tilde{\bm{S}}_n^{(k)}\!(\Bar{\bm{\epsilon}})\big\rangle\!\right)- \zeta(\bar{\bm{\epsilon}},\bar{\bar{\bm{\epsilon}}}).
    \end{multlined}
\end{align}
Noting that $\bm{\phi}_n\vert\bm{\phi}_{n-1},\bm{\Lambda}_a$ is Gaussian,
the message $\epsilon^{(\bm{\phi}_{n-1}\to \bm{\phi}_{n})}$ can be calculated as
\begin{multline}   
    \hspace{-4mm} \ln\big(\epsilon^{(\bm{\phi}_{n-1}\to \bm{\phi}_{n})}\big) \propto\\ -\frac{1}{2}\big\langle\bm{\phi}_n - \bm{T}\overline{\bm{\phi}}_{n-1}\big\vert \bm{G}^{-\top}\overline{\bm{\Lambda}}_a\bm{G}^{-1} \big\vert\bm{\phi}_n - \bm{T}\overline{\bm{\phi}}_{n-1}\big\rangle,
\end{multline}
leading to the following functional form for the message
\begin{align}
    \epsilon^{(\bm{\phi}_{n-1}\to \bm{\phi}_{n})} = \mathcal{N}\!\big(\bm{\phi}_n; \bm{T}\overline{\bm{\phi}}_{n-1}, \bm{G}^{-\top}\overline{\bm{\Lambda}}_a\bm{G}\big).
\end{align}
Similarly,
\begin{equation}
    \epsilon^{(\phi_{n+1}\rightarrow \phi_n)} = \mathcal{N}\!\big(\phi_n; \bm{T}^{-1}\bar{\phi}_{n+1},\bm{T}^{\top}\bm{G}^{-\top}\overline{\bm{\Lambda}}_a\bm{G}^{-1}\bm{T}\big).
\end{equation}

It is now possible to calculate the surrogate $q(\bm{\phi}_n)$ since all messages are Gaussians. The surrogate is a product of Gaussians which can be derived using the standard result
\begin{equation}\label{eq:combi_of_gaussians}
    \mathcal{N}(\boldsymbol{\mu}_{\text{total}},\boldsymbol{\Lambda}_{\text{total}}) = \prod_{n=0}^N \mathcal{N}(\boldsymbol{\mu}_n,\boldsymbol{\Lambda}_n),
\end{equation}
with
\begin{equation}\label{eq:combi_of_gaussians_2}
    \boldsymbol{\Lambda}_{\text{total}} = \sum_{n=0}^N \boldsymbol{\Lambda}_n, \phantom{mmm} \boldsymbol{\mu}_{\text{total}} = \boldsymbol{\Lambda}_{\text{total}}^{-1}\sum_{n=0}^N \boldsymbol{\Lambda}_n\boldsymbol{\mu}_n.
\end{equation}

We now derive the surrogate $q(\bm{\Lambda}_a)$. To simplify calculations we will impose a diagonal gamma prior on $\bm{\Lambda}_a$, i.e., $p(\bm{\Lambda}_{a,{i\neq j}}) = 0$, with shape parameter $\zeta/2$ and rate parameter $\chi/2$. With this, the log-likelihood can be expressed as,
\begin{multline}
    \hspace{-4mm} \ln(p(\bm{\phi}_n|\bm{\phi}_{n-1},\bm{\Lambda}_a)) \propto \frac{1}{2}\ln(|\bm{\Lambda}_a|) \\-\frac{1}{2}\big\langle \bm{G}^{-1}(\phi_n-\bm{T}\phi_{n-1})\big|\bm{\Lambda}_a\big|\bm{G}^{-1}(\bm{\phi}_n-\bm{T}\bm{\phi}_{n-1})\big\rangle
    \end{multline}
Taking the expectation w.r.t. $\bm{\phi}_n$ and $\bm{\phi}_{n-1}$ yields
\begin{multline}
    \hspace{-3.3mm} \mathbb{E}[\ln(p(\bm{\phi}_n|\bm{\phi}_{n-1},\bm{\Lambda}_{a}))] \propto \frac{1}{2}\sum_{k=1}^{4}\ln(|{\lambda}_{a,k}|)
    \\-\frac{1}{2}\sum_{k=1}^{4}\mathbb{V}_{n,n-1,k}\lambda_{a,k}    
    \end{multline}
with $\bm{\lambda}_a = \text{diag}(\bm{\Lambda}_a)$ and
\begin{multline}\label{eq:def_of_V}
    \hspace{-3.3mm} \mathbb{V}_{n,n-1,i} = \left\Vert\bm{G}^{-1}\left(\Bar{\bm{\phi}}_{n,i}-\bm{T}\bar{\bm{\phi}}_{n-1,i}\right)\right\Vert^2 \\+\bm{G}^{-1}\left(\bar{\bar{\bm{\phi}}}_{n,i,i}+ \bm{T}\bar{\bar{\bm{\phi}}}_{n-1,i,i}\bm{T}^\top\right)\bm{G}^{-\top},
    \end{multline}
which is a gamma distribution in $\lambda_i$. Hence, after carrying out the sum in \eqref{eq:surrogate_for_Lambda}, we obtain a gamma distribution with shape parameter $\alpha = (N+\zeta)/2$ and rate parameter $\beta = (\chi+\sum_{n=1}^N\mathbb{V}_{n,n-1})/2$. In this article, $\zeta = \chi = 1$.

With all messages computed, the complete algorithm can now be formulated as Alg.~\ref{al:algorithm}.

\begin{algorithm}[t]        
\caption{Multiple Radar Bayesian Localization and Tracking}\label{algo:MRBLaT}
\begin{algorithmic}[1]
\Procedure{MRBLaT}{$\bm{Z}_N^{(1)},\hdots,\bm{Z}_N^{(N_{\text{radar}})}$}

\Statex In parallel at each radar $k$:
\State $\left(\bar{\bm{\epsilon}}^{(\bm{Z}_N^{(k)}\rightarrow\bm{\phi}_N)},\bar{\bar{\bm{\epsilon}}}^{(\bm{Z}_N^{(k)}\rightarrow\bm{\phi}_N)}\right)\leftarrow \underset{\overline{{\bm{\epsilon}}},\overline{\overline{{\bm{\epsilon}}}}}{\text{argmin}}\,D_{KL}\!\left(\bm{Z}_N^{(k)}\right)$ using \eqref{eq:KL_div_abs}

\State Broadcast to all radars $\left(\bar{\bm{\epsilon}}^{(\bm{Z}_N^{(k)}\rightarrow\bm{\phi}_N)},\bar{\bar{\bm{\epsilon}}}^{(\bm{Z}_N^{(k)}\rightarrow\bm{\phi}_N)}\right)$ and save to memory

\Statex Local message passing at each radar:
\For{ite $\leftarrow 1 $ to $N_{\text{ite}}$}
    \For{$n\leftarrow 0\,\, \text{to}\,\, N$}
        \State $\bar{\bar{\bm{\phi}}}_n^{-1}\leftarrow \sum_{\epsilon \in \mathcal{N}_{\bm{\phi}_n}} \bar{\bar{\epsilon}}_n^{-1}$ 
        \State $\bar{\bm{\phi}}_n\leftarrow \bar{\bar{\bm{\phi}}}_n \sum_{\epsilon \in \mathcal{N}_{\bm{\phi}_n}}\bar{\bar{\epsilon}}^{-1}\bar{\epsilon}$
    \EndFor
    \If{$N>1$}
    \State $\alpha = N+1$
    \State $\bm{\beta} = \sum_{n=1}^N\bm{\mathbb{V}}_{n,n-1}$ using \eqref{eq:def_of_V}
    \State $\bar{\Lambda}_a^{-1} = \bm{\beta}/\alpha$
    \Else
    \State $\bm{\Lambda}_a \leftarrow \bm{\Lambda}_a^{(\text{init})}$
    \EndIf
\EndFor
\EndProcedure
\\
\Return $\left(\bar{\bm{\phi}}_0\hdots,\bar{\bm{\phi}}_N,\bar{\bar{\bm{\phi}}}_0,\hdots,\bar{\bar{\bm{\phi}}}_N\right)$
\end{algorithmic}
\label{al:algorithm}
\end{algorithm}
\section{Simulations}

\noindent We consider three $3\times3$ MIMO radars operating in TDM mode, transmitting complex baseband signals of duration $T_{T_x}$ with a linear chirp. The three radars are positioned \qty{50}{m} apart along the $x$-axis, with the first radar in Origo\footnote{Optimal radar placement is outside the scope of the paper.}. Parameter settings for the radars can be seen in Table~\ref{tab:params}. The data is generated using the model in \eqref{eq:freqdomain}.

\aboverulesep=0.3ex
\belowrulesep=0.3ex
\setlength\tabcolsep{1.1em}
\begin{table*}
    \centering
    \caption{Parameter settings}
    \begin{tabular}{ccccccccccc}
    \toprule
        $N_{T,R}$ & PRF          & $\rho$          & G & P            & $R_{\text{max}}$    & $f_c$         & BW            & $T_{T_x}$       & $f_s$          & $\sigma_w^2$ \\ \midrule
        3         & \qty{10}{Hz} & \qty{0.05}{m^2} & 1 & \qty{6.99}{W} & \qty{300}{m}  & \qty{10}{GHz} & \qty{20}{MHz} & \qty{16}{$\mu$ s} & \qty{256}{MHz} & $\text{BW}\cdot k_b \cdot 290$\\
    \bottomrule \\ \addlinespace[-0.2cm]
    \multicolumn{11}{l}{$k_b$ is the Boltzmann constant.}
    \end{tabular}
    \label{tab:params}
\end{table*}

\begin{figure}[b]
    \centering
    \includegraphics{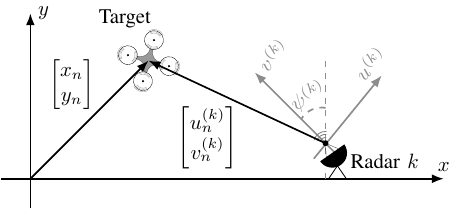}
    \caption{The relation between the local coordinate system $(u^{(k)},v^{(k)})$ and the global coordinate system $(x,y)$ which the algorithm runs in.}
    \label{fig:Coordinate_systems_fig}
\end{figure}

As a baseline, a KF using backwards smoothing was chosen, where the observation matrix was extended as follows,
\begin{equation}
    \bm{H} = \begin{bmatrix}
        \bm{H}_{\text{radar}}^{(1)} \\ \vdots \\\bm{H}_{\text{radar}}^{(N_{\text{radar}})}
    \end{bmatrix}, \phantom{mm} \bm{H}_{\text{radar}}^{(k)} = \begin{bmatrix}1 & 0 & 0 & 0 \\ 0 & 1 & 0 & 0 \end{bmatrix}\!.
\end{equation}
To have a fair comparison between the algorithms,  the KF also assumes linear motion for the target and the observations for the KF presupposes that a target exists. The range to the target can be found as follows,
\begin{equation}
    r_n^{(k)} = \text{median}\!\left(\underset{r}{\text{argmax}} \left|\mathcal{F}^{-1}\left\{\bm{Z}_n^{(k)}(r,\theta)\right\}\right|\right)\!,
\end{equation}
with $\mathcal{F}^{-1}\left\{\cdot\right\}$ being the inverse Fourier transform, and the median and argmax being taken over all $N_TN_R$ channels.
The direction of arrival $\theta^{(k)}_n$, is estimated using a Capon beamformer. The Cartesian coordinates for the radar under consideration are obtained as
\begin{equation}
    u_n^{(k)} = r_n^{(k)}\text{sin}\big(\theta_n^{(k)}\big),\phantom{mm} v_n^{(k)} = r_n^{(k)}\text{cos}\big(\theta_n^{(k)}\big).
\end{equation}
Each coordinate set $\big(u_n^{(k)},v_n^{(k)}\big)$ are now expressed in relation to the position and orientation of radar $k$. When combining the results, we translate these into a global coordinate system $\big(x_n^{(k)},y_n^{(k)}\big)$, as shown in Fig.~\ref{fig:Coordinate_systems_fig}. Denoting the position of radar $k$ in the global reference frame as $\bm{p}^{(k)}=[x_{\text{radar}}^{(k)}, y_{\text{radar}}^{(k)}]^\top$, the position may be translated into the same frame as
\begin{equation}\label{eq:transform_global}
    \begin{bmatrix}
        x_n^{(k)}\\y_n^{(k)}
    \end{bmatrix} = \bm{R}\left(\psi^{(k)}\right)\left(\begin{bmatrix}
        u^{(k)}_n \\v_n^{(k)}
    \end{bmatrix} -\bm{p}^{(k)}\right),
\end{equation}
where $\bm{R}$ is the rotation matrix evaluated in $\psi^{(k)}$ which is the angle between the second axis of the global coordinate system and the boresight of radar $k$. These global coordinates are then passed as inputs to the KF. 

When running MRBLaT, we minimize the KL divergence in \eqref{eq:KL_div_abs} numerically. We further simplify this task by assuming the covariance matrix $\bar{\bar{\bm{\epsilon}}}_n^{(k)}$ to be diagonal. For initialization of $N=0$ we use the same input as was given to the KF and for each subsequent iteration we use $\bm{T}\bar{\bm{\phi}}_{N-1}$ translated into the local coordinate system. Each radar uses the data to determine the statistics, i.e. $\big(\bar{\bm{\epsilon}}_n^{(k)},\bar{\bar{\bm{\epsilon}}}_n^{(k)}\big)$, and translates it back into the global frame using \eqref{eq:transform_global} which all radars run MRBLaT in.

\begin{figure}[t]
    \centering
    \includegraphics[width=1\linewidth]{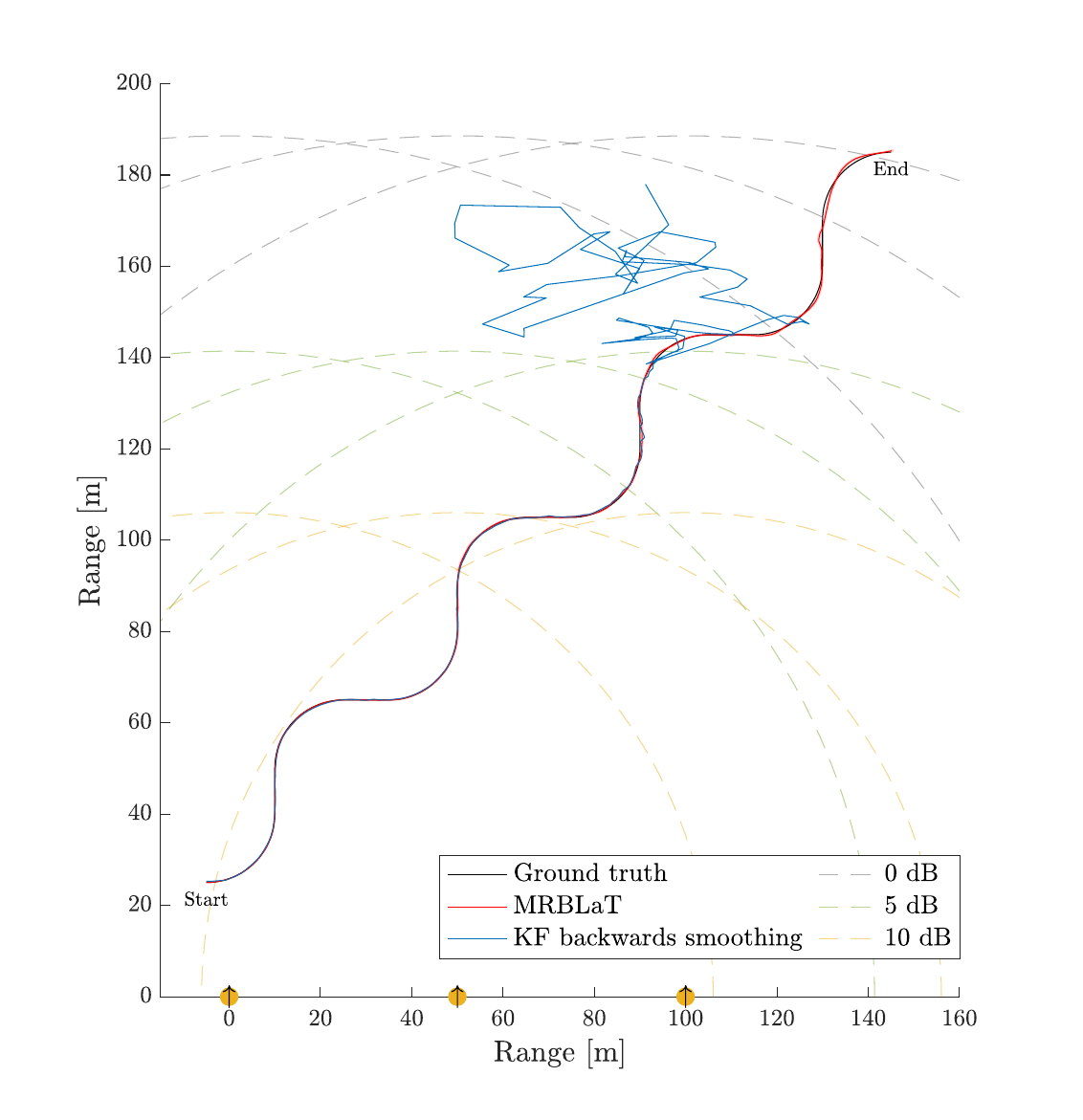}
    \vspace*{-10mm}    
    \caption{Track (A): Example of track estimates. The yellow dots represents the radars positions and the arrow represents the boresight direction. The red shaded area represents the \qty{95}{\percent} confidence interval for MRBLaT.}
    \label{fig:circmotion}
\end{figure}

\begin{figure}[t]
    \centering
    \includegraphics[width=1\linewidth]{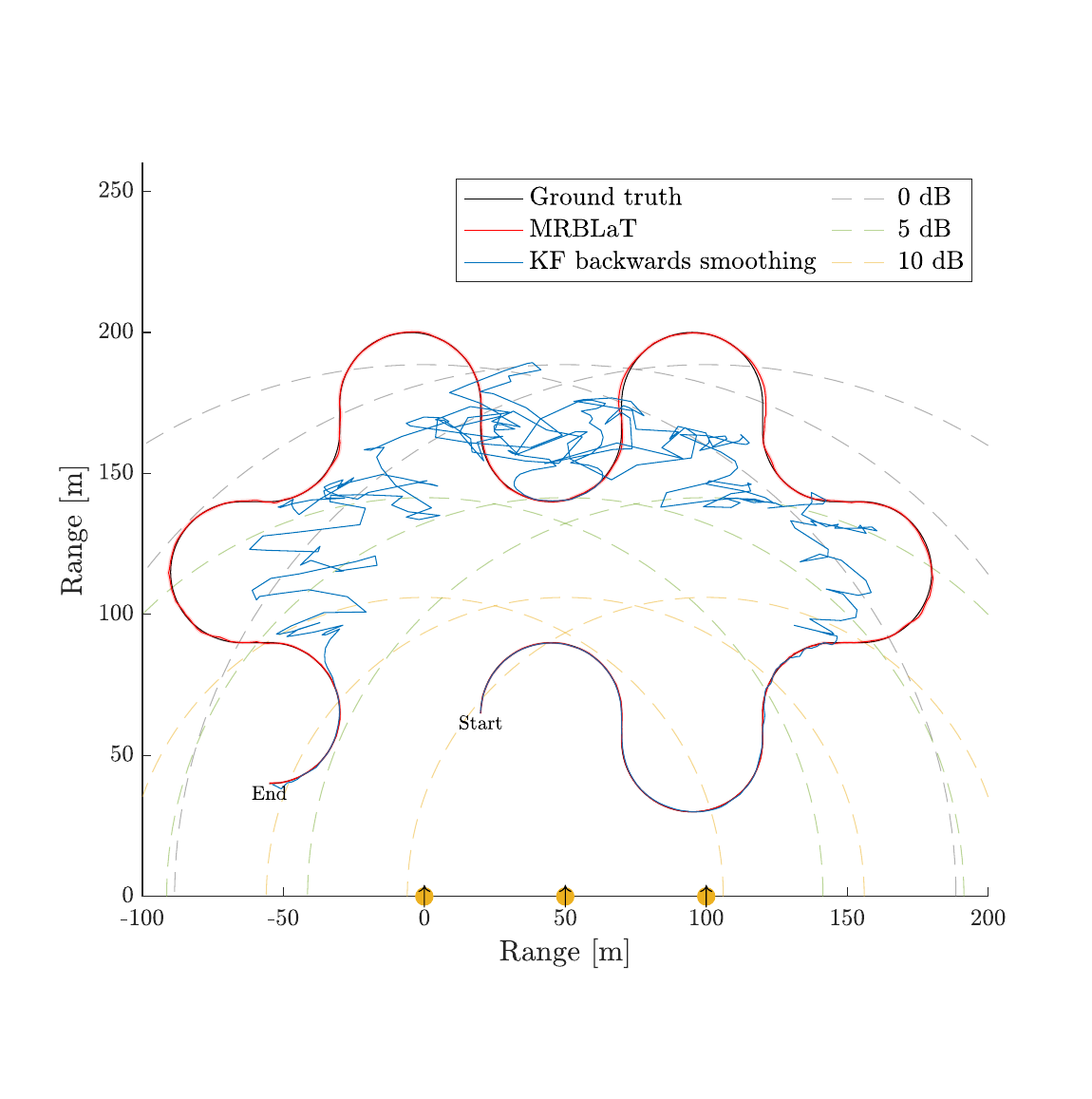}
    \vspace*{-15mm}
    \caption{Track (B): Example of track estimates. The yellow dots represents the radars positions and the arrow represents the boresight direction. The red shaded area represents the \qty{95}{\percent} confidence interval for MRBLaT.}
    \label{fig:S-shape}
\end{figure}

In the simulations, we evaluated tracks (A) and (B), shown in Fig.~\ref{fig:circmotion} and Fig.~\ref{fig:S-shape}, respectively. Along with the SNR contours, and 95\% point wise confidence intervals.
Both tracks consist of circle segments in differing directions at a constant velocity of \qty{10}{m/s}. When changing direction, the circle segments are separated by a linear motion where the drone comes to a full stop with a constant acceleration of \qty{10}{m/s^2}.

\begin{figure}
    \centering
    \includegraphics[width=1\linewidth]{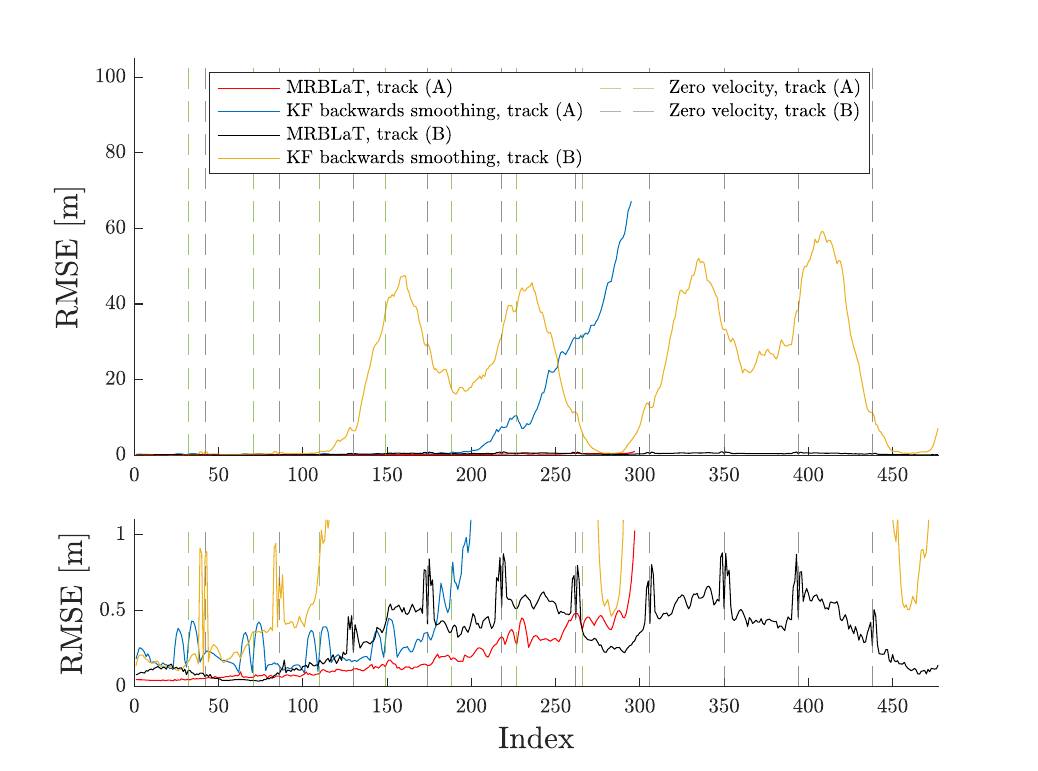}
    \caption{RMSE of track (A) and (B) calculated based on 512 Monte-Carlo simulations for two different $y$-axes.
    For track (A), the maximum RMSE is \qty{69}{m} for the KF and \qty{1.0}{m} for MRBLaT and for track (B), the maximum RMSE is \qty{59}{m} for the KF and \qty{0.9}{m} for MRBLaT. The vertical lines shows where the target has a velocity of zero.}
    \label{fig:RMSE}
\end{figure}

Figs.~\ref{fig:circmotion} and \ref{fig:S-shape} clearly shows that the MRBLaT algorithm outperforms the KF with backwards smoothing in tracking the target in low SNR conditions. Moreover, MRBLaT can track maneuvers deviating from the assumed kinematics. For track (A), the ground truth is covered by the confidence interval 84\% of the time. For track (B) this number is 93\%. 
When it does fall outside, the SNR is below \qty{1}{dB} suggesting the estimator underestimates the variance in this region.  
Contrary, the KF appears to completely lose track of the target when the SNR falls below \qty{5}{dB} for one or more radars as the variance of the estimates passed to the filter increases. 
Fig.~\ref{fig:RMSE} shows that the average RMSE for MRBLaT remains stable throughout the maneuvers for both tracks. For the KF, the RMSE of track (A) remains quite stable until index 200 where after it increases dramatically. For track (B), the RMSE of the KF is significantly larger than MRBLaT in index 110-270 and 280-450, which is in areas where the SNR is lower than \qty{5}{dB} for one or more radars.  
For both tracks, the RMSE generally fluctuates more when the target reaches zero velocity, but the impact is greater on the KF.

\section{Conclusion}
\noindent We propose a distributed localization and tracking algorithm with low communication overhead which scales linearly with the number of radars in the system. The algorithm builds on a Bayesian variational message passing framework, and outperforms a multi input KF in low SNR conditions. The algorithm tracks maneuvers differing from the assumed kinematics well across a wide range of SNR values, with the ground truth falling within the 95\% confidence interval of the estimator \qty{84}{\percent} and \qty{93}{\percent} of the time for track (A) and (B), respectively.

\bibliographystyle{IEEEtran}
\bibliography{litteratur}

\end{document}